\documentclass [preprint, prd, aps, nofootinbib]{revtex4}

\def\gtrsim{\begin{array}{c} > \\ \sim \end{array}}
\def\lesssim{\begin{array}{c} < \\ \sim \end{array}}

\usepackage{graphicx}

\input xy
\xyoption{all}

\begin{document}

\draft
\title{
Spacetime foam: from entropy and holography \\
to infinite statistics
and nonlocality}

\author{
Y. Jack Ng}

\address{Institute of Field Physics, Department
of Physics and Astronomy, University of North Carolina, Chapel
Hill, NC 27599, USA\\
E-mail: yjng@physics.unc.edu}
\bigskip

\begin{abstract}
Due to quantum fluctuations, spacetime is foamy on small scales.  The degree
of foaminess is found to be consistent with holography, a principle
prefigured in the physics of black hole entropy.  It has bearing on the
ultimate accuracies of clocks and measurements and the physics of quantum
computation.  Consistent with existing archived data on active galactic nuclei
from the Hubble Space Telescope, the application of the holographic 
spacetime foam model to cosmology
requires the existence of dark energy which, we argue, is composed of an
enormous number of inert ``particles" of extremely long wavelength.  We
suggest that these ``particles" obey infinite statistics in which all
representations of the particle permutation group can occur, and that the
nonlocality present in systems obeying infinite statistics may be related to
the nonlocality present in holographic theories.  We also propose to detect
spacetime foam by looking for halos in the images of distant quasars, and
argue that it does not modify the GZK cutoff in the ultra-high energy cosmic
ray spectrum and its contributions to time-of-flight differences of high
energy gamma rays from distant GRB are too small to be detectable.\\



\end{abstract}

\maketitle



\bigskip

\section{Introduction}

Following John Wheeler, many physicists (including the
author) believe that space is composed of an ever-changing arrangement of
bubbles called spacetime foam, also known as
quantum foam.  To understand the
terminology, let us consider the following simplified
analogy which Wheeler gave in a conference on gravity at University of
North
Carolina in 1957.  Imagine yourself flying an airplane
over an ocean.  At high altitude the ocean appears smooth.  But as you
descend, it begins to show roughness.  Close enough to the ocean surface,
you see bubbles and foam.  Analogously, spacetime appears smooth on a
large scale, but on sufficiently small scales, it will appear rough and
foamy, hence the term ``spacetime foam'' \cite{Wheeler}.
Many physicists believe
the foaminess is due to quantum fluctuations of spacetime, hence the
alternative term ``quantum foam.''  This reveiw article is devoted to a
discussion of spacetime foam models, or rather, a specific one of them
for which the concept of entropy plays a crucial role \cite{wigner,Karol,
found,PRL,llo04}.
The only ingredients we use in the whole discussion are quantum
mechanics and general relativity, the two pillars of modern
physics.  We hope that the results are very general and have wider
validity than most of the candidates of quantum gravity theory.
Assuming that there is a
unity of physics connecting the Planck scale to the cosmic
scale, we also apply quantum foam physics to
cosmology \cite{CNvD,Arzano,plb}.

In the next section, by a process of
mapping the geometry of a region of spacetime,
we show how foamy spacetime is.  It turns out that the degree of foaminess
of spacetime determines the maximum amount of
entropy a spatial region can hold.  This maximum amount of information can
be shown to be consistent with that encoded in the holographic principle
which has its origin in black hole physics.  Appropriately this spacetime
foam model has come to be known as the holographic model.  In section 3, we
propose to probe spacetime foam by looking for halos in the images of
distant quasars or bright active galactic nuclei (AGN). There we show that
the archive from the Hubble Space Telescople (HST) can already be used to
yield useful bounds.  Among the spacetime foam
models, the holographic foam model is unique in its correspondence with
the case of maximum energy density that a spatial region can hold without
collapsing into a black hole.  Applied to cosmology it ``predicts" a
critical energy density as observed in recent years.  This potential
connection between the extremely large and the extremely small
is explored in section 4; there we show that the
archive on quasars and AGN from HST can be used, in conjuction with the
physics of quantum
computation, to ``prove" the existence of dark energy, {\it independent}
of the various cosmological observations of recent years.  Indeed, from our
perspective, dark energy is arguably a cosmological manifestation of 
quantum foam!  In section 5,
we examine how and why dark energy is so different from ordinary
energy/matter.  There we show that the holographic model of spacetime foam
naturally leads us to speculate that dark energy consists of an incredibly
large number of extremely-long-wavelength ``particles". Then we exploit the
positivity of the entropy for these particles to show that they obey not the
ordinary bose or fermi statistics, but the exotic infinite statistics (also
known as quantum Boltzmann statistics).  We start our discussion with
holography and end with
infinite statistics.  We conclude section 5 by speculating that the
nonlocality known to be present in both of them may be related to each
other.  We give a summary in section 6.

For completeness, we discuss various other relevant topics of spacetime foam
physics in six short appendices.  In Appendix A, using a gedanken experiment
to measure distances we rederive the holographic foam model.  In Appendix B,
we apply the results in Appendix A to set bounds on the accuracies of clocks
and limits on computations, and derive the black hole entropy and lifetime.
We derive the holographic principle and the
Margolus-Levitin theorem (which is liberally used in the text) in Appendices
C
and D respectively.  Appendix E is devoted to a discussion of the
uncertainties
in energy-momentum measurements consistent with spacetime foam. In Appendix
F we apply the results obtained in Appendix E to discuss high-energy
$\gamma$ rays from distant gamma ray bursts and ultra-high energy cosmic ray
events.

On notations, the subscript ``P'' denotes Planck units; thus
$l_P \equiv (\hbar G /c^3)^{1/2} \sim 10^{-33}$ cm is the Planck
length etc.  On units, $k_B$ (the Boltzmann constant)
and sometimes $\hbar$ and c are put equal to 1 for simplicity.\\

\section{Quantum Fluctuations of Spacetime}

\subsection{Mapping the Geometry of Spacetime}

If spacetime indeed undergoes quantum
fluctuations, the fluctuations will show up when we measure a distance
$l$, in the form of uncertainties in the measurement.
 The question now is: how accurately
can
we measure this distance?  Let us denote by $\delta l$ the accuracy with
which we can measure $l$.
We will also refer to
$\delta l$ as the uncertainty or fluctuation of the distance $l$ for
reasons that will become obvious shortly.
%
One way to find $\delta l$ is to carry out a gedanken experiment to measure
$l$.  This is done in Appendix A.  But, for later use, it is
more convenient to find $\delta l$ by carrying out a process of mapping
the geometry of spacetime.
This method \cite{llo04,GLM}
relies on the fact that quantum fluctuations of spacetime
manifest themselves in the form of uncertainties in the geometry of
spacetime.  Hence the structure of spacetime foam can be inferred from the
accuracy with which we can measure that geometry.  Let us
consider mapping out the geometry of spacetime for a spherical volume of
radius $l$ over the amount of time $T = 2l/c$ it takes light to cross the
volume.  One way to do this is to fill the space with clocks, exchanging
signals with other clocks and measuring the signals' times of arrival.
This process of mapping the geometry of spacetime is a kind of computation,
in which distances are gauged by transmitting and processing information.
The total number of operations, including the ticks of the clocks and
the measurements of signals, is bounded by the Margolus-Levitin
theorem \cite{Lloyd}(see Appendix  D)
in quantum computation,
which stipulates that the rate of operations for any computer
cannot exceed the amount of energy $E$ that is available for computation
divided by $\pi \hbar/2$.   A total mass $M$ of clocks then
yields, via the Margolus-Levitin theorem, the bound on the total number of
operations given by $(2 M c^2 / \pi \hbar) \times 2 l/c$.  But to prevent
black hole formation, $M$ must be less than $l c^2 /2 G$.  Together, these
two limits imply that the total number of operations that can occur in a
spatial volume of radius $l$ for a time period $2 l/c$ is no greater than
$2 (l/l_P)^2 / \pi$.  To maximize spatial resolution, each clock must tick
only once during the entire time period.  The operations can be regarded as
partitioning the spacetime volume into "cells", then on the average each
cell
occupies a spatial volume no less than $(4 \pi l^3 / 3) / (2 l^2 /\pi l_P^2)
=
2 \pi^2 l l_P^2 /3$, yielding an average separation between neighhoring
cells
no less than $(2 \pi^2 /3)^{1/3} l^{1/3} l_P^{2/3}$.  This spatial
separation
is interpreted as the average minimum uncertainty \cite{hsu}
in the measurement of
a distance $l$, that is, \begin{equation}
\delta l \gtrsim l^{1/3} l_P^{2/3}, \label{vD2}
\end{equation}
where and henceforth (with a couple of exceptions) 
we drop multiplicative factors of order $1$.\\

\subsection{Models of Spacetime Foam}

We can now understand why this quantum foam model has come
to be known
as the holographic model.
Since, on the average, each cell occupies a spatial volume of $l l_P^2$,
a spatial region of size $l$ can contain no more than $l^3/(l l_P^2) =
(l/l_P)^2$ cells.  Thus this model
corresponds to the case of
maximum number of bits of information $l^2 /l_P^2$
in a spatial region of size $l$, that is
allowed by the holographic principle \cite{wbhts} (see AppendixC).

To see this more concretely, consider a cubic region of space
with linear dimension $l$.  
Conventional wisdom claims that the region can be partitioned into cubes
as small as $(l_P)^3$.  It follows that the number of degrees of freedom
of the region is bounded by $(l/l_P)^3$, i.e., the volume of the region in
Planck units.  But conventional wisdom is wrong, for according
to Eq. (\ref {vD2}), the smallest cubes into which we can partition the
region cannot have a linear dimension smaller than $(l l_P^2)^{1/3}$.
Therefore, the number of degrees of freedom of the region is bounded by
$[l/(l l_P^2)^{1/3}]^3$, i.e., the area of the region in Planck units,
as stipulated by the holographic principle \cite{wbhts}.

Recently Gambini and Pullin \cite{GPH} have shown that holography follows
from the framework of loop quantum gravity in spherical symmetry.  They have
derived from first principles an uncertainty in the
determination of volumes that grows radially (corresponding to the $l$
above); i.e., they have recovered Eq.~(\ref{vD2}).

It will prove to be useful to compare the holographic model in the mapping
of the geometry of spacetime
with the one that corresponds to spreading the spacetime cells uniformly
in both space and time.  For the latter case, each cell has
the size of $(l^2 l_P^2)^{1/4} =
l^{1/2} l_P^{1/2}$ both spatially and temporally so that each clock ticks
once in the time it takes to communicate with a neighboring clock.  Since
the dependence on $l^{1/2}$ is the hallmark of a random-walk fluctuation,
this quantum foam model corresponding to  $\delta l \gtrsim
(l l_P)^{1/2}$ is called the random-walk model \cite{AC}.
Compared to the holographic model, the random-walk model predicts a
coarser spatial resolution, i.e., a larger distance fluctuation,
in the mapping of spacetime geometry.  It
also yields a smaller bound on the information content in a spatial
region, viz., $(l/l_p)^2 / (l/l_P)^{1/2} = (l^2 / l_P^2)^{3/4} =
(l/l_P)^{3/2}$ bits.

Actually there are many other models of spacetime foam,
in addition to the holographic model and the random-walk model.
We can parametrize them according to $\delta l \sim l^{1 - \alpha}
l_P^{\alpha}$ for $\alpha$ of order 1, with $\alpha = 2/3$ and $1/2$ for the
holographic model and the
random-walk model respectively \cite{FordHu}.

Note that the minimum $\delta l$ just found for the
holographic model corresponds
to the case of maximum energy density $\rho = (3 / 8 \pi) (l l_P)^{-2}$ 
for a sphere of radius $l$ not
to collapse into a black hole.  Hence the holographic model, unlike
the other models, requires, for its
consistency, the energy density to have the critical value.
By contrast, for instance, the  random-walk model corresponds to an energy
density
$(l l_P)^{-2}\gtrsim \rho \gtrsim l^{-5/2} l_P^{-3/2}$.  (The upper bound
corresponds to the clocks ticking every $(l l_P)^{1/2}$ while the lower
bound corresponds to the clocks ticking only once during the entire time
$2l/c$.)\\

\subsection{Cumulative Effects of Spacetime Fluctuations}

Let us examine the cumulative effects \cite{NCvD}
of spacetime fluctuations over
a large distance.
Consider a distance $l$,
and divide it into $l/ \lambda$ equal
parts each of which has length $\lambda$ (that can be as small as $l_P$).
If we start with $\delta \lambda$ from each part, the question is how do the
$l/ \lambda$ parts
add up to $\delta l$ for the whole distance $l$.  In other words, we want
to find
the cumulative factor $C$ defined by
$\delta l = C\, \delta \lambda$.
Since $\delta l \sim l^{1/3} l_P^{2/3} = l_P
(l/l_P)^{1/3}$ and
$\delta \lambda \sim {\lambda}^{1/3} l_P^{2/3} =
l_P (\lambda/l_P)^{1/3}$, the result is
$C = (l /\lambda)^{1/3}$.  Note that
the cumulative factor is {\it not} linear in $(l/\lambda)$, i.e.,
$\frac{\delta l}{\delta \lambda} \neq \frac{l}{\lambda}$.  (In
general, it is
much smaller than $l/\lambda$).
The reason for this is obvious: the $\delta \lambda$'s (which take on $\pm$
sign with equal probability)
from the $l/ \lambda$
parts in $l$ do {\it not} add coherently.
In general, for spacetime foam models corresponding to $\delta l \sim l^{1 -
\alpha} l_P^{\alpha}$, the cumulative factor is given by $C = (l /
\lambda)^{1 - \alpha}$.  Thus for the random-walk model,
the cumulative factor is $(l/ \lambda)^{1/2}$.
Note that for the holographic case, the individual fluctuations cannot be
completely
random (as opposed to the random-walk model); strangely successive
fluctuations appear to be entangled and somewhat anti-correlated
(i.e., a plus fluctuation is slightly more likely followed by a minus
fluctuation and vice versa),
in order that together they produce a total fluctuation less
than that in the random-walk model.  This small amount of
anti-correlation between successive fluctuations (corresponding to
what statisticians call fractional Brownian motion with
self-similarity parameter $\frac{1}{3}$)
must be due to quantum gravity effects.

\begin{figure}[ht]

\centerline{\includegraphics[width=3.5in]{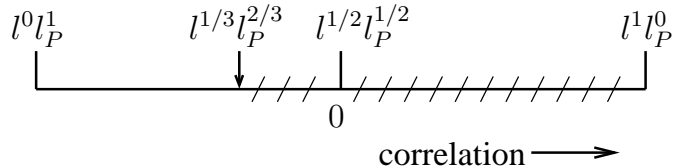}}
\caption{
Lower bounds on $\delta l$ for the various quantum gravity models.
The fluctuation of a distance $l$ is given by the
sum of $l/l_P$ fluctuations each by plus or minus $l_P$.
Spacetime foam appears to choose a small anti-correlation
(i.e., negative correlation) between
successive fluctuations, giving
a cube root dependence in the number
$l/l_p$ of fluctuations for the total fluctuation
of $l$ (indicated by the arrow).  The corresponding model falls between
the two extreme cases of complete randomness, i.e., zero
correlation (corresponding to $\delta l \sim l^{1/2}l_P^{1/2}$) and
complete anti-correlation (corresponding to $\delta l \sim l_P$).
\label{fig1}
}
\end{figure}

On the other hand, if
successive fluctuations are completely anti-correlated, then the fluctuation
of a distance $l$ is
given by the minuscule $l_P$,
independent of the size of the distance.
For completeness, we mention that {\it a priori} there are also models with
positive correlations between
successive fluctuations.  But these models yield unacceptably
large fluctuations in distance measurements ---
it turns out
(see next section) that these models (actually all models with $\alpha
\lesssim 0.6$
corresponding to the hatched line
in Fig.~\ref{fig1})
have already been observationally ruled out.\\

\section{Probing Quantum Foam with Extragalactic Sources}

The Planck length $l_P \sim 10^{-33}$ cm is so short that we need an
astronomical (even cosmological) distance $l$ for its fluctuation $\delta
l$ to be detectable.
Let us consider light (with wavelength $\lambda$)
from distant quasars or bright active galactic nuclei \cite{lie03,NCvD}.
Due to the quantum fluctuations of spacetime, the wavefront, while planar,
is itself ``foamy", having random fluctuations in phase \cite{NCvD} $\Delta
\phi \sim 2 \pi \delta l / \lambda$ and in
the direction of the wave vector \cite{CNvD} given by $\Delta \phi / 2 \pi
$ (for $\delta l \ll \lambda$).  In effect, spacetime foam creates a
``seeing disk" whose angular diameter is $\sim \Delta \phi /2 \pi  $.  For
an interferometer with baseline length $D$, this means that dispersion will
be seen as a spread in
the angular size of a distant point source, causing a reduction in the
fringe
visibility when $\Delta \phi / 2 \pi \sim \lambda / D  $. For a quasar of 1
Gpc away, at infrared wavelength, the
holographic model predicts a phase fluctuation $\Delta \phi \sim 2 \pi
\times
10^{-9}  $ radians.  On the other hand, an infrared interferometer (like the
Very Large Telescope Interferometer) with $D \sim 100$ meters has $\lambda /
D \sim 5 \times 10^{-9}  $.
Thus, in principle, this method will allow the use of interferometry fringe
patterns to test the holographic model!  Furthermore, these tests can be
carried out without guaranteed time using archived high resolution, deep
imaging data on quasars, and possibly, supernovae from existing and
upcoming telescopes.

The key issue here is the
sensitivity of the interferometer.  The lack of observed fringes may simply
be due to the lack of sufficient flux (or even just
effects originated from the turbulence of the Earth's atmosphere) rather
than the possibility that the
instrument has resolved a spacetime foam generated halo.
But, given sufficient sensitivity, the
VLTI, for example, with its
maximum baseline, presumably has sufficient resolution to detect spacetime
foam halos for low redshift quasars, and in principle, it can be even more
effective for the higher redshift quasars. Note that the test is simply a
question
of the detection or non-detection of fringes.  It is not a question of
mapping the structure of the predicted halo.

In the meantime, we can use existing archived data on
quasars or active galactic nuclei
from the Hubble Space Telescope to test the quantum foam models \cite{CNvD}.
Consider the case
of PKS1413+135 \cite{per02}, an AGN for which the redshift is $z = 0.2467$.
With $l \approx 1.2$ Gpc and $\lambda = 1.6 \mu$m,
we \cite{NCvD} find $\Delta \phi \sim 10 \times 2 \pi$ and
$10^{-9} \times 2 \pi$ for the random-walk model and
the holographic model of spacetime foam respectively.
With $D = 2.4$ m for HST, we expect to detect halos
if $\Delta \phi \sim 10^{-6} \times 2 \pi$.
Thus, the HST image only fails to test the holographic model by
3 orders of magnitude.

However, the absence of a quantum foam induced halo structure in the
HST image of PKS1413+135 rules out convincingly
the random-walk model.  (In fact, the scaling relation
discussed above indicates
that all spacetime foam models with $\alpha \lesssim 0.6$ are ruled
out by this HST observation.)  As we will see in the next section, this
result
has profound implications for cosmology  \cite{llo04,CNvD,Arzano,
plb}.\\

\section{From Quantum Foam to (Holographic Foam) Cosmology}

We can draw a useful conclusion \cite{llo04,CNvD,Arzano,plb}
from the {\it observed} cosmic critical density in the present era
(consistent with the prediction of the cosmology based on the holographic
model of spacetime foam, which henceforth
we call the holographic foam cosmology (HFC))
$\rho \sim H_0^2/G \sim (R_H l_P)^{-2}$ (about $10^{-9}$
joule per cubic meter),
where $H_0$ is the present Hubble parameter of the observable universe
and $R_H$ is the Hubble radius.
Treating the whole universe as a computer\cite{llo02, llo04}, one can
apply the Margolus-Levitin theorem to conclude that the universe
computes at a rate $\nu$ up to $\rho R_H^3 \sim R_H l_P^{-2}$
($\sim 10^{106}$ op/sec), for a total of $(R_H/l_P)^2$
($\sim10^{122}$) operations during its lifetime so far.
If all the information of this huge computer is stored in ordinary
matter, then we can apply standard methods of statistical mechanics
to find that the total number $I$ of bits is $[(R_H/l_P)^2]^{3/4} =
(R_H/l_P)^{3/2}$ ($\sim 10^{92}$).
It follows that each bit flips once in the amount of time given by
$I/\nu \sim (R_H l_P)^{1/2}$ ($\sim 10^{-14}$ sec).  On
the other hand, the average separation of neighboring bits is
$(R_H^3/I)^{1/3} \sim (R_H l_P)^{1/2}$ ($\sim 10^{-3}$
cm).  Hence, assuming only ordinary matter exists to store all the
information in the universe results in the conclusion that the time
to communicate with neighboring bits is equal to the time for each
bit to flip once.  It follows that the accuracy to which ordinary
matter maps out the geometry of spacetime corresponds exactly to
the case of events
spread out uniformly in space and time discussed above for the case
of the random-walk model of spacetime foam.

But, as shown in the previous section,
the sharp images of distant quasars or active galactic nuclei
observed at the Hubble Space
Telescope have ruled out the random-walk model. From the demise of the
random-walk model and the fact that ordinary matter only contains an
amount of information dense enough to map out spacetime at a level
consistent with the random-walk model, one can now infer that
spacetime must be mapped to a finer spatial accuracy than that which
is possible with the use of ordinary matter.  But if ordinary matter
does not do, there must be another kind of substance with which spacetime
can be mapped to the observed accuracy, conceivably as given
by the holographic model. The natural conclusion \cite{CNvD,Arzano,plb} is
that unconventional
(dark \cite{Turner}) energy/matter exists!
Note that this argument does not make use of
the evidence from recent cosmological (supernovae, cosmic microwave
background, and galaxy clusters)
observations \cite{SNa}.

A comparison between what the random-walk model and the holographic model
yield for the entropy bound etc. is given in Table 1.  See the next section
for the explanation of the last column.

\begin{table}
Table 1.  Random-walk model versus holographic model.
The corresponding quantities for the random-walk model (second
row)
and the holographic model (third row) of spacetime foam (STF) appear
in the same columns in the following Table.  (Entropy is measured
in Planck units.)\\

\centering
\begin{tabular}{|c|c|c|c|c|c|} \hline
STF & distance &   entropy & energy & matter/ & type of \\
model & fluctuations &  bound & density & energy & statistics \\  \hline
\hline
random- & $\delta l \gtrsim l^{1/2} l_P^{1/2}$ &   $(Area)^{3/4}$ & $(l
l_P)^{-2} \gtrsim \rho$ & ordinary  & bose /  \\ walk & & & $ \gtrsim
l^{-5/2} l_P^{-3/2} $ & & fermi \\  \hline \hline
holo- & $\delta l \gtrsim l^{1/3} l_P^{2/3}$ &   $Area$ & $\rho \lesssim (l
l_P)^{-2}$ & dark & infinite \\
graphic & & & & energy & \\  \hline
\end{tabular}
\end{table}

The fact that our
universe is observed to be at or very close to its critical density
must be taken as solid albeit indirect evidence in favor of the
holographic model because, as discussed above, it is the only
model that requires, for its consistency,
the maximum energy density without causing gravitational collapse.
Specifically, according
to the HFC, the cosmic density is
\begin{equation}
\rho = (3 /8 \pi) (R_H l_P)^{-2} \sim (H/l_P)^2,
\label{density}
\end{equation}
and the cosmic entropy is given by \begin{equation}
I \sim (R_H /l_P)^2.
\label{bits}
\end{equation} Thus the average energy carried by each bit is
$\rho R_H^3/I \sim R_H^{-1}$ ($\sim 10^{-31}$ eV).  Such
long-wavelength (hence ``non-local'')
bits or ``particles'' carry negligible kinetic energy.
Also according to HFC, it takes each unconventional bit
the amount of time
$I/\nu \sim R_H$ to flip.  Thus, on the average, each bit flips
once over the course of cosmic history.  Compared to the conventional
bits carried by ordinary matter, these bits are rather passive and
inert (which, by the way, may explain why dark energy is dark). This is
understandable since each unconventional bit has,
at its disposal, only
such a minuscule amount of energy.  But together (there can be as many as
$(R_H/l_P)^2 \sim 10^{123} $ of them in the present observable universe, far
outnumbering the $ 10^{92}$ or so particles of ordinary matter and
radiation), they supply the
missing mass/energy of the universe.  Accelerating the cosmic expansion is
a relatively simple task, computationally speaking.\\

\section{Infinite Statistics and Nonlocality}

\subsection{Infinite Statistics}


What is the overriding difference between conventional matter and
unconventional energy/matter (i.e., dark energy and perhaps also dark
matter)?
To find that out, let us first
consider a perfect gas of $N$ particles obeying Boltzmann statistics
(which, rigorously speaking, is not a physical statistics but is still a
useful statistics to work with) at temperature $T$ in a volume $V$.  For the
problem at hand, as the lowest-order approximation, we can neglect the
contributions from matter and radiation to the cosmic energy density for the
recent and present eras.  Then it can be
shown that the Friedmann equations for $\rho \sim H^2 /G$ are solved by $H
\propto 1/a$ and $a \propto t$,
where $a(t)$ is the cosmic scale factor.
Thus let us take $V \sim R_H^3$, $T \sim R_H^{-1}$, and $N \sim
(R_H/ l_P)^2$. A standard calculation (for the relativistic case) yields the
partition function $Z_N = (N!)^{-1} (V / \lambda^3)^N$, where
$\lambda = (\pi)^{2/3} /T$.
With the free energy given by
$F = -T ln Z_N = -N T [ ln (V/ N \lambda^3) + 1]$,
we get, for the entropy of the system,
\begin{equation}
S = - ( \partial F / \partial T)_{V,N} = N [ln (V / N \lambda^3) + 5/2].
\label{entropy1}
\end{equation}
For the non-relativistic case with the
effective mass $m \sim R_H^{-1}$ (coming from some sort of potential with
which we are not going to concern ourselves),
the only changes in the above expressions are given by the substitution
$\lambda \longrightarrow (2 \pi / mT)^{1/2}$.
With $m \sim T \sim R_H^{-1}$, there is no significant
qualitative difference between the non-relativistic and
relativistic cases.

The important point to note is that, since $V \sim \lambda^3$, the entropy
$S$ in Eq. (\ref{entropy1}) becomes nonsensically negative unless $ N \sim
1$ which is equally nonsensical because $N$ should not be too different from
$(R_H/l_P)^2 \gg 1$.
Intentionally we have calculated the entropy by employing the familiar
Boltzmann statistics (with the correct Boltzmann counting factor),
only to arrive at a contradictory result. But now the solution to this
contradiction
is pretty obvious: the $N$ inside the log in Eq. (\ref{entropy1}) somehow
must be absent.  Then $ S \sim N
\sim (R_H/l_P)^2$ without $N$ being small (of order 1) and S is non-negative
as physically required.  That is the case if the ``particles" are
distinguishable and nonidentical!  For in that case, the Gibbs $1/N!$ factor
is absent from the partition function $Z_N$, and the entropy becomes
\begin{equation}
S = N[ln (V/ \lambda^3) + 3/2].
\label{entropy2}
\end{equation}

Now the only known consistent statistics in greater than two space
dimensions
without the Gibbs factor (recall that the Fermi statistics and Bose
statistics give
similar results as
the conventional Boltzmann statistics at high temperature)
is infinite statistics (sometimes called
``quantum Boltzmann statistics") \cite{DHR,govorkov,greenberg}.  Thus we
have
shown that the ``particles" constituting dark energy obey infinite
statistics,
instead of the familiar Fermi or Bose statistics \cite{plb}.
What is infinite statistics?  Succinctly, a Fock realization of infinite
statistics is provided by a $q$ deformation of the commutation relations of
the oscillators:
$a_k a^{\dagger}_l - q a^{\dagger}_l a_k = \delta_{kl}$ with $q$ between -1
and 1 (the case $q = \pm 1$ corresponds to bosons or fermions).  States are
built by acting on a vacuum which is annihilated by $a_k$.  Two states
obtained by acting with the $N$ oscillators in different orders are
orthogonal.  It follows that the states may be in any representation
of the permutation group.  The statistical mechanics of particles obeying
infinite statistics can be obtained in a way similar to Boltzmann
statistics, with the crucial difference that the Gibbs
$1/N!$ factor is absent for the former.  Infinite statistics can be
thought of as corresponding to the statistics of identical particles with an
infinite number of internal degrees of freedom, which is
equivalent to the statistics of nonidentical particles since they are
distinguishable by their internal states.\\

\subsection{Nonlocality}

Infinite statistics appears to have one ``defect": a theory of particles
obeying infinite statistics cannot be local \cite{fredenhagen,
greenberg,arzanobala}. (That is, the fields associated with infinite
statistics are not local,
neither in the sense that their observables commute at spacelike separation
nor in the sense that their observables are pointlike functionals of the
fields.)  The expression for the number operator (for the case of $q = 0$)
\begin{equation}
n_i = a_i^{\dagger} a_i + \sum_k a_k^{\dagger} a_i^{\dagger} a_i a_k +
\sum_l \sum_k a_l^{\dagger} a_k^{\dagger} a_i^{\dagger} a_i a_k a_l + ...,
\label{number}
\end{equation}
is both nonlocal and nonpolynomial in the field operators,
and so is the Hamiltonian.  The lack of
locality may make it difficult to formulate a relativistic verion of the
theory; but it appears that a non-relativistic theory can be developed.
Lacking locality also means that the familiar spin-statistics relation is no
longer valid for particles obeying infinite statistics; hence
they can have any spin.  Remarkably, the TCP theorem and cluster
decomposition have been shown to hold despite the lack of locality
\cite{greenberg}.

Actually the lack of locality for theories of infinite statistics may have a
silver lining.  According to the holographic principle, the
number of degrees of freedom in a region of space is bounded not by
the volume but by the surrounding surface.  This suggests that the
physical degrees of freedom are not independent but, considered
at the Planck scale, they must be infinitely correlated, with the result
that the spacetime location of an event may lose its invariant significance.
Since the holographic principle is believed to be
an important ingredient in the formulation of quantum gravity,
the lack of locality for theories of infinite statistics may not be a
defect; it can actually be a virtue.  Perhaps it is this lack of
locality that makes it easier to incorporate gravitational interactions in
the theory.  Quantum gravity and infinite statistics appear to fit together
nicely, and nonlocality seems to be a common feature of both of them
\cite{plb}.

We note the following related work.  Using the Matrix theory approach,
Jejjala, Kavic and Minic \cite{minic} have argued that dark energy
quanta obey infinite statistics (and
that the fine structure of dark energy is governed by
a Wien distribution).  They have also concluded that the non-locality
present in systems obeying infinite statistics and the non-locality
present in holographic theories may be related.

Strominger \cite{stromvolo} has shown that the wave
function of many similarly charged extremal
black holes depends on each black
hole's position, and thus the black holes can be considered as
distinguishable and accordingly obey infinite statistics.

More generally Giddings \cite{SG} has observed that the non-perturbative
dynamics of gravity is nonlocal.  His argument is
based on several reasons: lack of a precise definition in quantum
gravity --connected with the apparent absence of local observables;
indications from high-energy gravitational scattering; hints
from string theory, particularly the AdS/CFT correspondence;
conundrums of quantum cosmology; and the black hole information paradox.
Horowitz \cite{GH} has noted that quantum gravity may need
some violation of locality; thus when one reconstructs
the string theory from the gauge theory (in the AdS/CFT correspondence),
physics may not be local on all length scales.
Meanwhile Ahluwalia \cite{Ahl} has argued that when measurement
processes involve energies of the order of the Planck scale, the
fundamental assumption of locality may no longer be a good approximation.
He has shown that position measurements alter the
spacetime metric in a fundamental manner and this unavoidable
change in the spacetime metric destroys the commutativity (and
hence locality) of position measurement operators.\\

\section{Summary, Discussion and Conclusion}

\begin{figure}
\[
\xymatrix{
& *\txt{QM $+$\\ bh physics} \ar[d] \ar[dr]^(.6){\txt{bh entropy}}  & \\
\Omega \cong 1 \ar@{<->}[r] \ar[d]_{\txt{AGN data}} &
*\txt{holographic\\ st foam} \ar@{<->}[r] \ar[d] & *\txt{holography}
\ar@{--}[dl]^(.4){\txt{nonlocality}} \\ \txt{dark energy} \ar[r] &
\infty~\textrm{statistics} & }
\]
\caption{Connecting the different ideas.}
\label{figure2}
\end{figure}
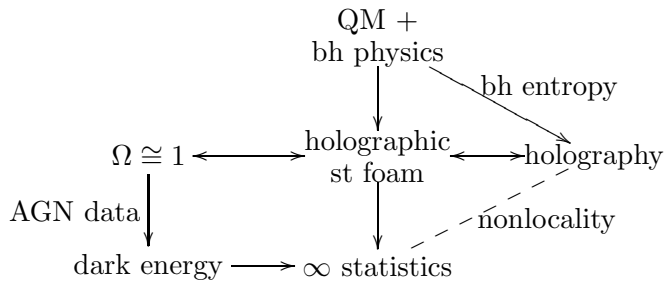

A short summary of this review article is given by the accompanying
flow chart Figure~\ref{figure2}.

Starting from quantum mechanics and some rudimentary black hole physics
(viz, there is an upper bound of matter/energy that can
be put in a spatial region without the region collapsing into a black hole),
we derive the holographic model of spacetime foam. From the simple
observation that the cosmic energy density is critical (i.e., the density
parameter of the universe $\Omega \cong 1$),
aided by some archived data on quasar or AGN from the Hubble Space
Telescope, we are led to conclude that dark energy exists.
This observation is a solid piece of evidence
in favor of the holographic foam model which, after all, ``predicts''
that the cosmic energy density has the critical value. One is then invited
to apply the holographic model to cosmology, leading to the logical
speculation that the constituents of dark energy obey infinite
statistics.  We have now come full circle, starting with holography
and ending with infinite statistics, with both sharing one common
property: nonlocality.  In the whole discussion, entropy plays a
leading role, first in motivating holography, and secondly in
determining, via its positivity property, that the ``particles''
constituting dark energy obey infinite statistics.

We conclude with some observation and a partial list of open questions.

(1) We have considered a perfect gas consisting of ``particles" of extremely
long wavelength,
obeying Boltzmann statistics in the Universe at temperature $T$.  But we
have
seen that these ``particles" have had interactions only of order one time on
the
average during the entire cosmic history.
A question can be raised as to whether such an inert gas can come to thermal
equilibrium at any well defined temperature.  We do not have
a good answer; but the fact that all these ``particles", though
extremely inert, have a wavelength comparable to the observable Hubble
radius may
mean that they overlap significantly, and accordingly can perhaps share a
common temperature.

(2) These ``particles'' provide a (more or less)
spatially uniform energy density, like a time-dependent cosmological
constant \cite{ChenWu}.  But
in a way, this type of models is preferrable to the cosmological constant
because it may be easier to understand a zero cosmological constant (perhaps
due to
a certain not-yet-understood symmetry or initial condition \cite{unimod})
than an
exceedingly small (but non-zero) cosmological constant.

(3) Recall that earlier cosmic epochs are associated with $\rho \propto
a^{-4}$ (radiation-domination) and (followed by)
$\rho \propto a^{-3}$ (matter-domination).  If the holographic foam
cosmology
is correct, these epochs are succeeded by the dark-energy-dominated
era with $\rho \propto a^{-2}$.  We note that the successes of the
conventional big bang cosmology, such as the primordial nucleosynthesis,
are not affected by this form of dark energy.

(4) An obvious question concerns the sign of the pressure for the gas of
dark energy ``particles''.
For cosmic energy density $\rho \sim H^2/G$, the equation of state is
given by $w = p/\rho \sim -1/3$, not negative enough to give an
accelerating expansion.  However, it has been pointed out \cite{negativep}
that a transition from
an earlier decelerating to a recent and present
accelerating cosmic expansion can arise as a pure interaction
phenomenon if pressureless dark matter is coupled to holographic
dark energy \cite{holocos}.
As a bonus, within the framework of such cosmological models,
we can now understand
why, in addition to dark energy, dark matter has to exist.
On the other hand, this scenario will become less natural if the equation of
state $w$ turns out to be very close to $-1$.

(5) Critical cosmic energy density is the hallmark of the inflationary
paradigm \cite{zizzi}.  Can
the holographic foam cosmology (which requires $\Omega \cong 1$)
supplement inflation in solving cosmological problems and in providing the
necessary primordial
perturbations to yield the observed astronomical structures?  What are the
phenomenological consequences of HFC?

(6) While it
appears quite reasonable that holography (and quantum gravity in general)
and infinite statistics are compatible, exactly how they are related is not
yet clear.  In particular, how is the nonlocality
in holography related to the nonlocality present in theories with
infinite statistics?  (We venture a guess: Entropy is the common link, so it
may hold the key in understanding the relation.)

Before last century, spacetime was regarded as nothing
more than a passive and static arena in which events took place.
Early last century, Einstein's general relativity
changed that viewpoint and promoted spacetime to an
active and dynamical
entity.  Quantum mechanics blossomed around that time.  But
the challenge to understand the quantum nature of spacetime
was not taken up seriously until quite a bit later.
Now many physicists believe
that spacetime, like all matter and energy, undergoes quantum
fluctuations.  These quantum fluctuations make spacetime
foamy on small spacetime scales.
In this article,
we have used a global positioning-like system to measure
the geometry of spacetime,
to show that spacetime fluctuations, on the average, scale as
the cube root of distances.  This result is very general, depending
only on quantum mechanics and limited black hole physics (viz., the size of
a black hole scales linearly with its mass).
The cube root dependence is
strange, but it has been shown to be consistent with the
holographic principle and (as shown in Appendix B) semi-classical black hole
physics in general.  Furthermore, applied to cosmology, it successfully
``predicts" the
existence of dark energy and that the cosmic energy density is critical.  To
the author at least, the cube-root result for spacetime foam
is as beautiful as it is strange --- and, when the Very Large
Telescope Interferometer reaches its design performance, it may even be
proven to be true.\\

\bigskip

\section*{Acknowledgments}
This work was supported in part by the US Department of Energy and the
Bahnson Fund of the University of North Carolina.  I thank H.~van Dam,
G.~Amelino-Camelia, W.~Christiansen,
T.~W. Kephart, O.~W. Greenberg, M.~Arzano,
J.~Pullin, D.~Minic, L.~Mersini, R.~Rohm, X.~Calmet, S.~Hsu, T.~Biswas,
N.~Mavromatos, S.~Giddings, V.~Husain, D.~Pavon, and W.~Zimdahl
for useful discussions, L.~L. Ng for his help in the preparation of this
manuscript, and P.~Zizzi for his invitation to write this review article.

\bigskip

{\bf Appendix A: Salecker \& Wigner's Gedanken Experiment}

Let us conduct a thought experiment to measure
a distance $l$.
Following Salecker \& Wigner \cite{SW} we can put a clock (of mass $m$)
at one end of the distance and a mirror
at the other end.  By sending a light signal from the clock to the mirror
in a timing experiment, we can determine the distance $l$. But the clock's
position jiggles according to Heisenberg's uncertainty principle in quantum
mechanics, resulting in an uncertainty in the measurement:
\begin{equation}
\delta l^2 \gtrsim \frac{\hbar l}{mc}.
\label{sw}
\end{equation}
The jiggling of
the clock is reduced by using a massive clock.  But, according to general
relativity, a massive clock would distort the surrounding space severely,
affecting adversely the accuracy in the measurement of the distance:
\begin{equation}
\delta l \gtrsim \frac{Gm}{c^2}.
\label{ngvan}
\end{equation}

The
conflicting requirements from quantum mechanics and general relativity
reach a compromise, resulting in the elimination of the dependence on
$m$, and yielding \cite{wigner,found,PRL} (by taking
the product
of Eq.~(\ref{ngvan}) and Eq.~(\ref{sw})):
\begin{equation}
\delta l \gtrsim (l l_P^2)^{1/3} = l_P \left(\frac{l}{l_P}\right)^{1/3}.
\label{nvd1}
\end{equation}
Obviously the accuracy of distance measurements is intrinsically limited by
this
amount of uncertainty or fluctuation.\\

{\bf Appendix B: From Space-time Fluctuations to Black Holes}

To gain confidence in the strange scaling of space-time fluctuations with
the cube-root of distances, let us look for theoretical ``evidence''.
Fortunately such circumstantial evidence does exist --- in the sector of
black hole physics.  To
show that, we have to make a small detour to consider clocks and
computers \cite{Barrow,PRL} first.\\

{\it Clocks}

Consider a clock (of mass $m$), capable
of resolving time to an accuracy of $t$, for a period of time
$T$.
Then bounds on the resolution time and the running time of the clock can be
derived by following an argument very similar
to that used above in the analysis of the gedanken experiment to measure
distances.
Actually, the two arguments are so similar that one can identify the
corresponding quantities.  [See Table 2.]

\begin{table}
Table 2.  Distance measurements, clocks and computers.
The corresponding quantities in the discussion of distance
measurements (first column), clocks
(second column), and computers (third column) appear
in the same rows in the following Table.\\

\centering
\begin{tabular}{||c||c||c||} \hline \hline
distance & clocks & computers \\
measurements & & \\   \hline\hline
$\delta l/c$ & $t$ & $1/\nu$\\
\hline
$l/c$ & $T$ & $I/\nu$\\  \hline
$\delta l^2 \gtrsim \hbar l/mc$ & $t^2 \gtrsim \hbar T / mc^2$
& $I \nu \lesssim mc^2/\hbar$  \\
\hline
$\delta l \gtrsim Gm/c^2$ & $t \gtrsim Gm/c^3$
& $\nu \lesssim c^3/Gm$  \\
\hline
$l/(\delta l)^3 \lesssim l_P^{-2}$
($\delta l \gtrsim l^{1/3} l_P^{2/3}$)
& $T/t^3 \lesssim t_P^{-2}$
& $I \nu^2 \lesssim t_P^{-2} = c^5/\hbar G$        \\ \hline \hline
\end{tabular}
\end{table}

Following the argument in the previous Appendix, we obtain
\begin{equation}
t^2 \gtrsim \frac{\hbar T}{mc^2},\hspace{.5in}
t \gtrsim \frac{Gm}{c^3},\hspace{.5in}
t \gtrsim t_P \left(\frac{T}{t_P}\right)^{1/3},
\label{clock1}
\end{equation}
the analogs of Eq.~(\ref{sw}), Eq.~(\ref{ngvan}) and
Eq.~(\ref{nvd1}) respectively.  The last inequality yields a fundamental
limit on how accurate a clock can be in measuring a given amount of
time $T$.  Recently Gambini, Porto and Pullin \cite{GPP} have applied this
result to argue that, since one cannot have a perfectly classical clock
in nature, quantum mechanics needs to be modified in that quantum
evolution is not unitary.  Such a non-unitary evolution leads to
potentially observable decoherence, e.g., in a system of Bose-Einstein
condensates.  These authors have further claimed that, in
real life, one could never observe the black hole information
paradox, since quantum states decohere (due to our lack of perfect
clocks) at a rate faster than the one an evaporating black hole makes it
disappear.\\

{\it Computers}

We can easily translate
the above relations for clocks into useful relations for a simple
computer.
Since the resolution
time $t$ for clocks is the smallest time interval relevant in the
problem, the fastest
possible processing frequency is given by its reciprocal, i.e., $1/t$.
Thus if $\nu$
denotes the clock rate of the computer, i.e., the number of operations
per bit per unit
time, then it is natural to identify $\nu$ with $1/t$.
To identify the number $I$
of bits of information in the memory space of a simple computer, we recall
that the running time $T$ is the longest time interval relevant in the
problem.  Thus,
the maximum number of steps of information processing is given by the
running time divided by the resolution time, i.e., $T/t$.
It follows that one can identify the number $I$ of bits of the
computer with $T/t$.  (One can think of a tape of length $cT$
as the memory space, partitioned into bits each of length $ct$.)
The bounds on the precision and lifetime of a clock
given by Eq.~(\ref{clock1}) are
now translated into bounds on the rate of computation and number of bits
in the computer, yielding respectively
\begin{equation}
I \nu \lesssim \frac{mc^2}{\hbar},\hspace{.6in}
\nu \lesssim \frac{c^3}{Gm}, \hspace{.6in} I \nu^2 \lesssim \frac{c^5}{\hbar
G}.
\label{computer}
\end{equation}
The first inequality shows that the speed of computation is bounded by
the energy of the computer divided by $\hbar$,
in agreement with the Margolus-Levitin theorem (see Appendix D).\\

{\it Black Holes}

Now we can apply what we have learned about clocks and computers
to black holes of mass $m$ \cite{Barrow,PRL}.
Let us consider using a black hole to measure time.
It is reasonable to use
the light travel time around the black hole's horizon as the resolution
time of the clock,
i.e., $t \sim Gm/c^3 \equiv t_{BH}$, then
from the first equation in Eq.~(\ref{clock1}), one immediately finds that
\begin{equation}
T \sim \frac{G^2 m^3}{\hbar c^4} \equiv T_{BH},
\label{Hawking}
\end{equation}
which is the celebrated
Hawking black hole lifetime.

Finally, let us consider using a black hole to do computations
\cite{PRL,llo02,llo04,Barrow}.  This may
sound like a ridiculous proposition.  But if we believe that black holes
evolve according to quantum mechanical laws, it is
possible, at least in
principle, to program black holes to perform computations that
can be read out of the fluctuations in the Hawking black hole
radiation.
How large is the memory space of a black hole computer, and
how fast can it compute?
Applying the results for computation derived above, we readily find
the number of bits in the memory space of a black hole computer, given by
the lifetime of the black hole divided by its resolution time as a clock,
to be
\begin{equation}
I = \frac{T_{BH}}{t_{BH}} \sim \frac{m^2}{m_P^2} \sim \frac{r_S^2}{l_P^2},
\label{bhcomputer1}
\end{equation}
where $m_P = \hbar/(t_P c^2)$ is the Planck mass, and $r_S^2$ denotes the
event horizon area of the black hole.
This gives the number $I$ of bits as the event horizon area in Planck units,
in agreement with
the identification of black hole entropy.
(Note that
Eq.~(\ref{bhcomputer1}) can also be derived from Eq.~(\ref{computer}).)
Furthermore, the number of operations per unit time for a
black hole computer is given by
$I \nu = (T_{BH} / t_{BH}) \times (1 / t_{BH})
\sim mc^2 / \hbar$, viz.,
its energy divided by Planck's constant.
(It is interesting that all the bounds on clocks and computation
discussed above are actually saturated by black hole clocks and black hole
computers respectively.)  Thus it is
now abundantly clear that the graininess of space-time due to quantum
fluctuations that we found above is entirely consistent with well-known
black hole physics.

We conclude this Appendix with a speculation on the nature of black hole
microstates. \cite{microstates}
It is baffling that ordinary matter configurations on the
verge of becoming black holes have entropy $ S \sim (Area)^{3/4}$
whereas the black holes themselves have entropy $S \sim Area$.  What can
account for this difference in the amount of entropy?  Inspired by the
columns under the headings of ``entropy bound" and ``type of statistics"
in Table 1, we make the following conjecture.  While ordinary matter
obeys bose or fermi statistics, it is possible that the microscopic
constituents of black holes {\it effectively} obey infinite statistics.  To
wit, to a black hole of
size $r_S$ and Hawking temperature $\sim 1/r_S$, the application of
Eq. (\ref{entropy2}) readily yields $ S \sim (r_S / l_P)^2$  by
taking the number of microscopic constituents to be $N \sim (r_S / l_P)^2$.
(On the other hand, we do not see how Eq. (\ref{entropy1}) can give
an entropy that scales as the area $r_S^2$.)
That said, it is hard to show how matter obeying bose or fermi
statistics can become constituents obeying infinite statistics
upon gravitational collapse to form a black hole.\\

{\bf Appendix C: Holographic Principle}

In essence, the holographic principle\cite{wbhts}
stipulates that although the world
around us appears to have three spatial dimensions, its contents can
actually be encoded on a two-dimensional surface, like a hologram.
In other words, the maximum entropy of a region of space is given (aside
from
multiplicative factors of order $1$ which we ignore as we have so far)
by its surface area in Planck units.  This result can be derived
by appealing to black hole physics and the second law of theromodynamics
as follows \cite{wbhts}.
Consider a system with entropy $S_0$ inside a spherical
region $\Gamma$ bounded by surface area $A$.  Its mass must be less
than that of a black hole with horizon area $A$ (otherwise it would have
collapsed into a black hole).  Now imagine a spherically
symmetric shell of matter collapsing onto the original system with just the
right amount of energy so that together with the original mass, it forms
a black hole which just fills the region $\Gamma$.  The black hole so formed
has entropy $S \sim A/l_P^2$.  But according to the second law of
thermodynamics, $S_0 \leq S$.  It follows immediately that $S_0
\lesssim A/l_P^2$, and hence the maximum entropy of a region of space is
bounded by its surface area, as asserted by
the holographic principle.\\

{\bf Appendix D: the Margolus-Levitin Theorem}

The Margolus-Levitin theorem \cite {Lloyd} plays an important role
in our discussion in the text.  For the sake of completeness,
in this Appendix we follow Margolus and Levitin to derive it.
Consider the maximum speed of dynamical evolution for a given
physical system.  Assume
that the system has a discrete energy spectrum $E_n, n = 0, 1,2 ...$ with
the lowest
energy chosen to be $E_0 = 0$.  We can write an arbitrary state
$|\psi \rangle$ as a superposition of energy eigenstates, with coefficients
$c_n$ at time $t = 0$.  Let $|\psi_0 \rangle$ evolve for a time $t$ to
become
$|\psi_t \rangle$.  Denote the transition amplitude $\langle \psi_0 | \psi_t
\rangle$ by
$S(t)$.  We want to find the smallest value of $t$ such that $S(t) = 0$.  To
do that, we note that
$Re(S) = \sum_n |c_n|^2 cos (E_n t / \hbar)$.
Using the inequality $cos x \geq 1 - (2/ \pi)(x + sin x)$, valid for $x \geq
0$, we get
$Re(S) \geq 1 - 2Et/(\pi \hbar) + 2 Im(S) / \pi$, where $E$ denotes the
average energy of the system.
But $S(t) = 0$ implies both $Re(S) = 0$ and $Im(S) = 0$.  So this inequality
becomes
$0 \geq 1 - 4Et/ h$, where $ h = 2 \pi \hbar$.
Thus the earliest that $S(t)$ can possibly equal zero is when $ t = h/4E$.
Applied to a computer, this result implies that
the maximum speed of computation is given by $4/h$ times the
energy available for computation.\\

{\bf Appendix E: Energy-Momentum Fluctuations}

Just as there are uncertainties in spacetime measurements, there are
also uncertainties in energy-momentum measurements due to
spacetime foam effects.  Thus there is a limit to how accurately we
can measure and know the energy and momentum of
a system \cite{wigner}.
Imagine sending
a particle of momentum $p$ to probe a certain structure of spatial extent
$l$ so that $p \sim \hbar/l$.
It follows that $\delta p \sim (\hbar/l^2) \delta l$.  Spacetime
fluctuations
$\delta l \gtrsim l (l_P/l)^{2/3}$ can
now be used to give
\begin{equation}
\delta p = \beta p \left(\frac{p}{m_P c}\right)^{2/3},\hspace{.8cm}
\delta E = \gamma E \left(\frac{E}{E_P}\right)^{2/3},
\label{dp}
\end{equation}
where {\it a priori} $\beta \sim 1$ and $\gamma \sim 1$,
$E_P = m_P c^2 \sim 10^{19}$ GeV is the Planck energy and we have added
the corresponding statement for energy uncertainties.
We emphasize that all the uncertainties take on $\pm$ sign with equal
probability (most likely, something like
a Gaussian distribution about zero).\\

{\it Modified Dispersion Relations}

Energy-momentum uncertainties affect both the energy-momentum conservation
laws and dispersion relations \cite{Ng}.  Energy-momentum is
conserved up to energy-momentum uncertainties due to quantum foam effects,
i.e., $\Sigma (p_i^{\mu} + \delta p_i^{\mu}$)
is conserved, with $p_i^{\mu}$ being the average values
of the various energy-momenta.  On the other hand
the dispersion relation is now generalized to read
\begin{equation}
E^2 - p^2c^2 - \epsilon p^2c^2 \left({pc \over E_P}\right)^{2/3} = m^2c^4,
\label{moddisp}
\end{equation}
for high energies with $E \gg mc^2$.  {\it A priori} we expect
$\epsilon \sim 1$ and is independent of $\beta$ and $\gamma$.  But since
the dispersion relation is actually derived from $(E + \delta E)^2 -(p +
\delta p)^2c^2 =
m^2c^4$, with $\delta p$ and $\delta E$ given by Eqs.~(\ref{dp}), the
coefficients in Eqs~(\ref{dp}) and (\ref{moddisp}) are related as
\begin{equation}
\epsilon = 2( \beta - \gamma).\\
\label{related}
\end{equation}

{\it Fluctuating Speed of Light}

The modified dispersion relation discussed above has an interesting
consequence for the speed of light \cite{ACetal}.
Applying Eq.~(\ref{moddisp})
to the massless photon yields
its speed \cite{Ng}
\begin{equation}
v = \frac{\partial E}{\partial p} \simeq c
\left( 1 +\frac{5}{6} \epsilon
\frac{E^{2/3}}{E_P^{2/3}}\right),
\label{gams}
\end{equation}
which is energy-dependent and fluctuates around c.
For example,
a photon of $10^{13}$eV energy has a speed
fluctuating about $c$ by about 1 cm/sec. \\

{\it Unmodified Threshold Energies in Collisions}

Consider the scattering process in which an energetic
particle
of mass $m_1$ collides head-on with a soft
photon of
energy $\omega$ in the production of two energetic particles with
mass
$m_2$, $m_3$, as in the interaction of an ultra-high energy
cosmic particle with the Cosmic Microwave Background Radiation.
After taking into account energy-momentum uncertainties,
energy-momentum conservation, and the modified dispersion relations
Eq. (\ref{moddisp}),
we obtain the threshold energy equation \cite{Ng}
\begin{equation}
E_{th} = p_0 + \tilde{\eta} {1\over 4\omega}
   {E_{th}^{8/3} \over E_P^{2/3}},
\label{appen5}
\end{equation}
where
$p_0 \equiv [(m_2 + m_3)^2 - m_1^2]/4 \omega$ is the (ordinary) threshold
energy if there are no energy-momentum
uncertainties, and
$\tilde{\eta} \equiv \eta_1 - [\eta_2 m_2^{5/3} +
        \eta_3 m_3^{5/3}]/(m_2 + m_3)^{5/3}$,
with
$\eta_i \equiv 2\beta_i - 2\gamma_i - \epsilon_i$. With the aid of
Eq.~(\ref{related}), we obtain $\eta_i = 0$.  It follows that
the threshold energies are {\it not} modified. Now recall that, a priori,
quantum fluctuations can {\it lower} as well
as raise the reaction thresholds.  But the lowering of reaction thresholds
can give rise to matter instability which is not observed, as pointed out by
Aloisio et al. \cite{Aloisio}.  With the threshold energies not modified,
the serious problem of matter instability is avoided.\\

{\bf Appendix F: Gamma Ray \& Cosmic Ray Phenomenologies}

{\it High Energy $\gamma$ Rays from Distant GRB}

Recall that, due to quantum fluctuations of spacetime, the
speed of light fluctuates around $c$ and the fluctuations increase
with energy ($\delta v \sim c (E/E_P)^{2/3}$, according to
Eq.~(\ref{gams})).  Thus
for photons emitted simultaneously from a distant
source, we expect an energy-dependent
spread in their arrival times.  To maximize the spread in arrival
times, we should look for energetic photons from distant sources.
So the idea is to look for a noticeable spread in arrival
times for high energy gamma rays from distant gamma ray
bursts.  This proposal was first made by G. Amelino-Camelia
{\it et al.} \cite{ACetal} in another context.
But the time-of-flight differences $\delta t$
increase only with the cube root of the
average overall time $t$ of travel ($\delta t \sim t^{1/3} t_P^{2/3}$)
from the gamma ray bursts to our
detector, leading to a time spread too small to be detectable \cite{Ng}.
Thus, if the spread in
arrival times for the photons recently observed \cite{MAGIC}
by the MAGIC gamma-ray telescope during a flare of the active galaxy
Markarian 501 is indeed confirmed to be due to quantum gravity effects
\cite{Ellisetal}, then
these effects are beyond those associated with quantum foam discussed
in this article. \\

{\it Ultra-High Energy Cosmic Ray Events}

Theoretically one expects the
UHECRs to interact with the Cosmic Microwave Background
Radiation and produce pions.
These interactions above the threshold energy
should make observations of UHECRs
with $E > 6 {\cdot} 10^{19}$eV (the GZK limit)
unlikely.  Since the reaction threshold energies are not modified, we
expect {\it no violation} of the GZK bound due to quantum foam effects.
(However this bound does not apply to cosmic rays coming from nearby sources
\cite{Auger}.)






\end{document}